\documentstyle[11pt,paspconf,epsf]{article}

\begin{document}

\title{GHRS Observations of the Lyman $\alpha$ Forest}
\author{Simon L. Morris}
\affil{Dominion Astrophysical Observatory, National Research Council,
       5071 West Saanich Road, Victoria, B.C., V8X 4M6, Canada, 
       e-mail: simon@dao.nrc.ca.}

\begin{abstract}

I review the results obtained using the GHRS on low redshift Lyman
$\alpha$ absorbers. Until the advent of HST and the GHRS, the existence
of such absorbers was doubted. The confirmation of their existence, in
one of the first GHRS GTO team results to be published, must rank as
one of the HSTs most interesting results.  The GHRS resolution allows
us to probe equivalent widths well below those detectable with the FOS,
and has led to a number of interesting new questions. One example is
the apparent disagreement between the GHRS result that there are many
Lyman $\alpha$ absorbers which are {\bf not} associated with luminous
galaxies, and FOS studies which suggest that {\bf all} such absorbers
have a nearby galaxy causing them. This almost certainly shows that the
equivalent width (or column density) range reachable by the GHRS
includes gas from a wide range of causes, and not only the halos of
luminous galaxies. With these data, we are seeing the debris left over
from Galaxy formation, material flung out from galaxy interactions and
starbursts, and also diffuse halo material at the outer edges of normal
galaxies.

\end{abstract}


\section{The picture prior to HST} \label{sec-prior}

\begin{figure}[ht]
\plotfiddle{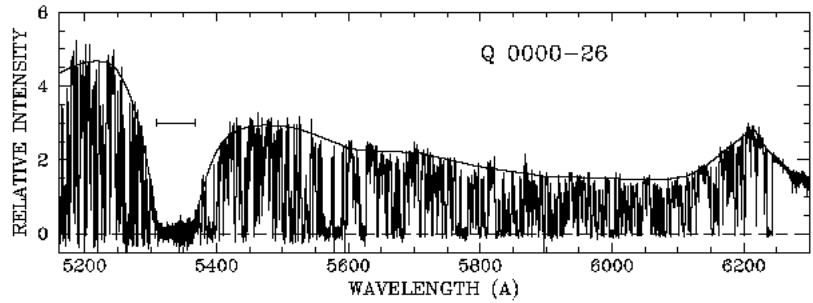}{5.5cm}{270}{55}{55}{-216}{216}
\caption{The high redshift Lyman $\alpha$ forest, taken from Lu et al. 1996. 
This is a Keck HIRES spectrum of Q0000-26, in arbitrary flux
units, with resolution $\sim$6.6 km~s$^{-1}$. Only the Lyman $\alpha$
forest portion is shown in order to illustrate the adopted continuum
level (smooth solid curve). Note the damped Lyman $\alpha$ absorption
at z$_{abs}$=3.39 or $\lambda$=5337{\AA}.
\label{fig-morriss1.eps}}
\end{figure}

The Lyman $\alpha$ `forest' refers to the dense overlapping absorption
seen shortward of the Lyman $\alpha$ emission line in high redshift QSO
spectra (see figure~\ref{fig-morriss1.eps} for an example, taken from
\cite{lu96}).  
This absorption is caused by a huge number of foreground condensations
of warm gas, with a significant component of neutral hydrogen. Nice
reviews of the observational situation just prior to the launch of HST
are given by \cite{carswell88}, and \cite{hunstead88}. A non-specialist
summary would be: the detectable neutral hydrogen column densities in
the forest vary from less than 10$^{13}$~cm$^{-2}$ up to
$\sim$10$^{17}$~cm$^{-2}$, above which the systems become optically
thick to the Lyman continuum (referred to as Lyman-limit systems), and
generally show Mg~II absorption. There is a roughly power law fall-off
in density of absorbers with column density, such that it is considered
reasonable to associate these rarer Lyman-limit systems with the inner
regions of the halos of luminous (L*) galaxies. Above
10$^{15.5}$~cm$^{-2}$, C~IV absorption is often detected, and so these
systems are called `metal-line' systems, although recent Keck HIRES
observations suggest that the carbon abundances are comparable in the
lower column density systems as well (around 0.01 solar, \cite{cowie95}
and \cite{songaila96}). The absorbers have line widths corresponding to
temperatures in the range 10$^{4-5}$ K, consistent with their being
photo-ionised by the UV background, and do not show strong clustering.
All the above led to a model in which the absorption was produced by
inter-galactic clouds, probably made up of pristine material left over
from (or not yet part of) the process of galaxy formation. It has also
been noted (see for example \cite{rauch95a}) that estimates of the
ionisation fraction in these clouds, along with big-bang limits for the
baryon density of the universe, suggest that, at high redshift, these
low column density absorbers may contain a substantial fraction of {\bf
all} the baryons in the universe. The Lyman $\alpha$ forest at high
redshift is not an exotic trace component.

\begin{figure}[ht]
\plotfiddle{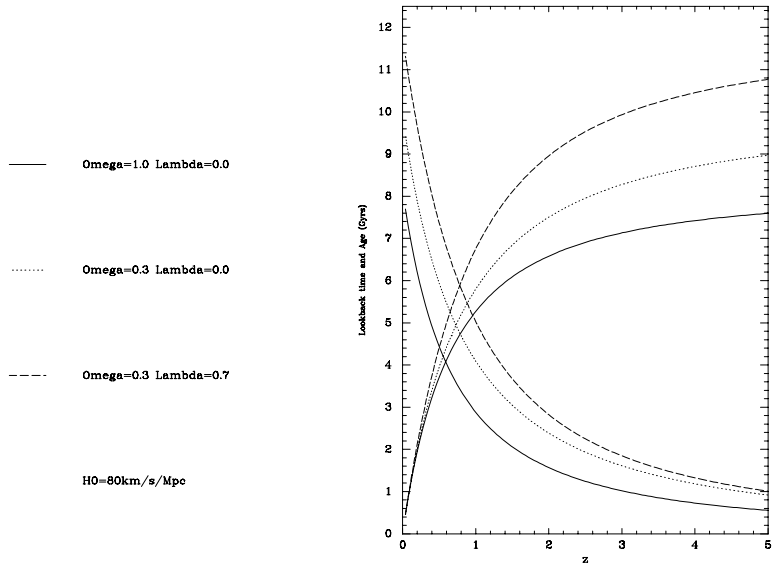}{9cm}{90}{50}{50}{180}{-36}
\caption{Lookback time and age of the Universe in units of 10$^9$ years
for 3 different cosmologies. The main point is to note that the
redshift range of Lyman $\alpha$ absorbers visible from the ground
(1.6$<$z$<$5) only covers ages from around 1 to 3 Gyrs. 
\label{fig-morriss2.eps}}
\end{figure}

The biggest problem in understanding the forest prior to launch of the
Hubble Space Telescope (HST) was that Lyman $\alpha$ absorption could
only be seen from the ground when it had been redshifted beyond
3200{\AA}, i.e. a redshift of 1.6. As can be seen from
figure~\ref{fig-morriss2.eps}, this only covers ages from around 1 to 3
Gyrs,
leaving the period from 3 Gyrs to the present (8-12 Gyrs) unobservable.
It also meant that redshift surveys for galaxies at the same redshifts
as the absorbers were impossible, and so tests of the association
between absorbers and galaxies could not be made. 

However, over the redshift range from z=4 to z=1.6, a steep fall-off in
absorber density per unit redshift was seen (see \cite{lu91} for
example). Fitting a power law to this fall-off, and extrapolating to
the present day, one would predict that there should be essentially no
detectable Lyman $\alpha$ absorbers left, making the advantages of UV
studies of the forest moot.

\section{HST/GHRS observations of 3C273.0 and line densities} \label{sec-3C273}

After the launch of HST, one of the Goddard High Resolution
Spectrograph (GHRS) team's highest priority projects was to observe the
brightest QSO in the sky -- 3C273.0. This QSO, with a redshift of
0.158, has an optical magnitude of 12.8, and IUE observations had
confirmed its strong UV flux, making it the best extragalactic target
for absorption line studies.

GHRS observations a low resolution with the G140L grating, and medium
resolution with the G160M grating were obtained, and were described in
\cite{morris91}. Low resolution FOS observations were published at
the same time by \cite{bahcall91}.
\begin{figure}[ht]
\plotfiddle{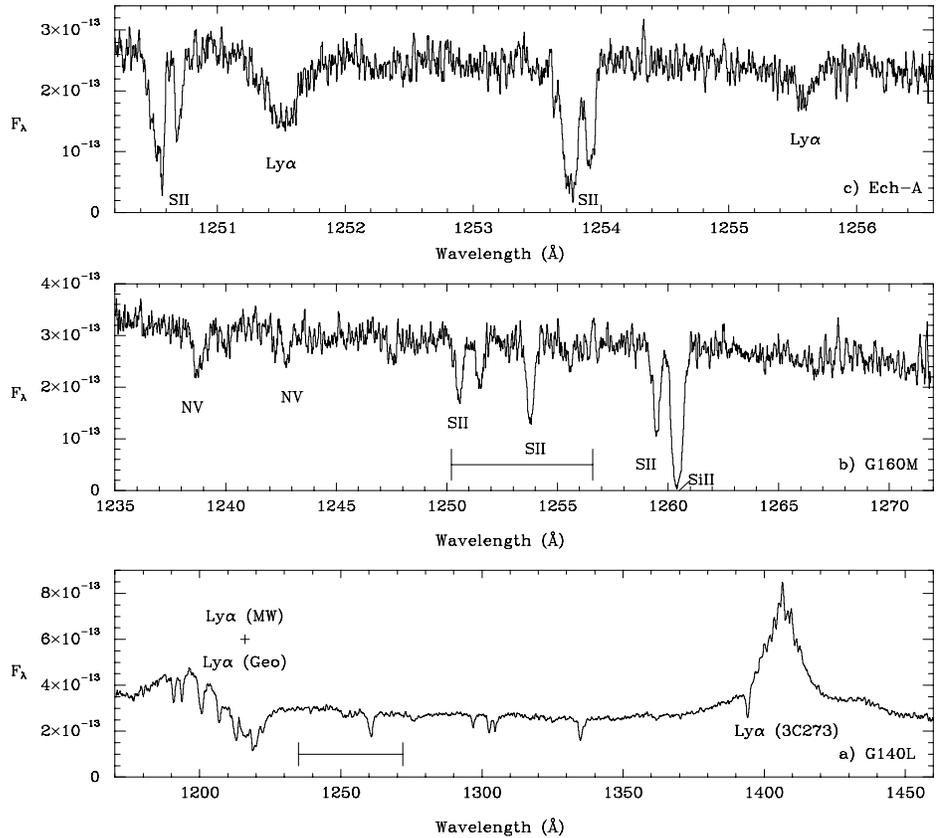}{11cm}{90}{70}{70}{288}{-54}
\caption{GHRS Spectra of the QSO 3C273.0. (a) GHRS G140L spectrum with
resolution of $\sim$2,000, (b) GHRS G160M spectrum with resolution
$\sim$15,000, (c) GHRS Ech-A spectrum with resolution $\sim$100,000.
All the spectra have been smoothed with a boxcar of width 4 pixels,
matching the spectral resolution. \label{fig-morriss3.eps}}
\end{figure}
Figure~\ref{fig-morriss3.eps}a is the GHRS G140L spectrum, with
resolution of $\sim$2,000, showing the entire wavelength range over
which Lyman $\alpha$ absorption can be seen.  Marked around 1250{\AA}
is the region observed with the G160M grating shown in
Figure~\ref{fig-morriss3.eps}b. This spectrum has resolution
$\sim$15,000, and shows a large number of strong galactic absorption
lines (discussed in detail in \cite{savage93}). Features which can
not be ascribed to ions in our galaxy are assumed to be Lyman $\alpha$
absorption from clouds along the line-of-sight (LOS) to 3C273.0. Marked
around 1253{\AA} is one of the regions observed with the Ech-A grating.
Figure~\ref{fig-morriss3.eps}c shows this (as yet unpublished) GHRS
Ech-A spectrum, which was obtained much later, with resolution
$\sim$100,000.  In it, the strong Lyman $\alpha$ line near 1251.5{\AA}
can be seen to be clearly resolved. With a resolution of $\sim$3
km~s$^{-1}$, this is one of the highest dispersion observations of an
extra-galactic Lyman $\alpha$ absorption at {\bf any} redshift.  The
original GHRS data set was published in \cite{brandt93}, while an
additional G160M spectrum of the interesting low redshift region
covering the Virgo cluster velocity range was published by 
\cite{weymann95}.  The three panels of Figure~\ref{fig-morriss3.eps} show
how valuable high spectral resolution is in detecting and studying weak
absorption lines, such as those of the Lyman $\alpha$ forest.

The studies above showed that Lyman $\alpha$ absorption was indeed
still present at the present day, and hence that some flattening of the
fall-off in line density with time must occur. Since then, the HST
Absorption Line Key Project, carried out with the FOS, has observed many
more LOS, albeit at low resolution, and has begun to quantify the
change in slope of the number density with redshift for the stronger
lines (see \cite{bahcall93}, \cite{bahcall96}, with a dissenting view presented by 
\cite{bechtold94}). Thus, although the luxuriant Lyman $\alpha$ forest at
high redshift has been dramatically thinned out, there remains a Lyman
$\alpha$ savannah at the present day, suitable for study.

\section{Correlations between absorbers and galaxies} \label{sec-correlations}

\begin{figure}[ht]
\plotfiddle{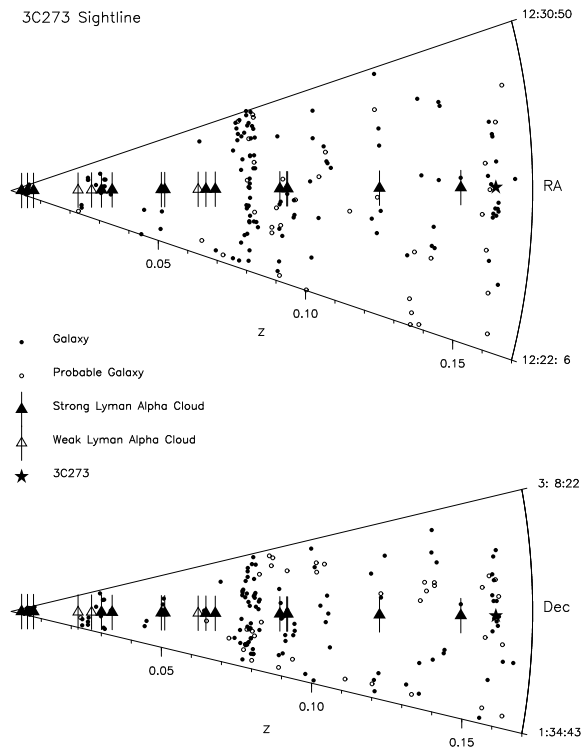}{12cm}{0}{70}{70}{-144}{0}
\caption{Pie-diagrams for the galaxies and absorbers observed in the
3C273.0 LOS from Morris et al. 1993. \label{fig-morriss4.eps}}
\end{figure}

At last Lyman $\alpha$ absorbers, with column densities in the same
range as those making up the forest at high redshift, were available at
distances within which galaxies could be studied. So, are the absorbers
always associated with a luminous galaxy, or are they inter-galactic
clouds randomly distributed in space? In the 3C273.0 LOS, as many as 17
absorbers were identified. \cite{morris93} published a galaxy
redshift survey complete to an optical magnitude of $\sim$19, and out
to a radius of 1$\deg$ from the QSO LOS.  Figure~\ref{fig-morriss4.eps}
shows the locations of absorbers and galaxies from that survey.
Angles have been exaggerated by a factor 15 to prevent overcrowding of
the symbols, which results in a highly distorted plot.  Initially
spherical structures (such as the 3C273.0 cluster of galaxies) appear
elongated transverse to the line of sight. \cite{morris93} also
published deep optical imaging around the QSO, searching for faint
dwarf galaxies near the LOS, which was followed up by \cite{rauch95b}. 
Neither of these studies found any suitable candidates. \cite{morris93}
showed narrow band imaging, searching for Balmer line
(H$\alpha$) emission at the same velocity as a couple of the lowest
redshift absorbers. No such emission was found. This line emission was
looked for with higher sensitivity by \cite{williams93}, who
claimed a detection, but this was shown not to be real by
\cite{bland94}. Finally, \cite{morris93} and \cite{vangorkom93} searched 
for 21~cm radio emission, without
success.

\begin{figure}[ht]
\plotfiddle{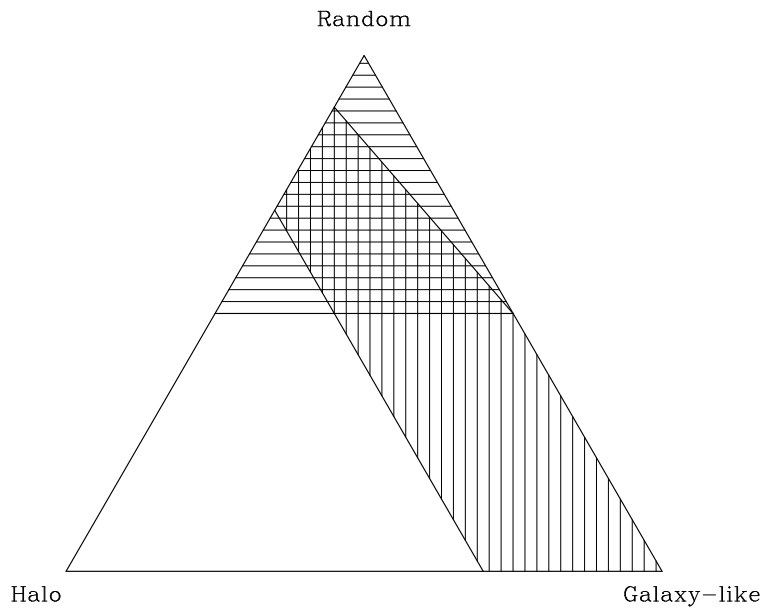}{9cm}{270}{60}{60}{-252}{324}
\caption{Schematic diagram showing the permitted combinations of the 3
Lyman $\alpha$ model populations from Mo and Morris 1994. Vertical
shading is the region permitted by the 0.5 Mpc scale correlations,
while horizontal shading shows the region permitted by the 10 Mpc scale
correlations. \label{fig-morriss5.eps}}
\end{figure}

After this intensive search of the 3c273.0 LOS, there remained clear
cases of absorbers with no galaxy of any sort within a radius of 4-5
Mpc.  On the other hand, statistical analysis of the correlation
between the absorbers and galaxies by \cite{morris93}, and examined in
more detail by \cite{mo94}, showed that the absorbers definitely were
not distributed at random with respect to the galaxies. These results
are summarised schematically in Figure~\ref{fig-morriss5.eps}.
Three Lyman $\alpha$ model populations were considered by \cite{mo94}: 
randomly distributed, distributed like galaxies, but not
actually part of an observed galaxies halo, or in fact part of an
observed galaxy's halo. The vertices of the triangle represent models
with a single type of absorber (random, galaxy-like or halo).  Moving
along the sides of the triangle represents different admixtures of two
different types of absorber, while the inner regions represent mixtures
of all three types of model. Marked along the edges are the permitted
ranges found by the Monte-Carlo tests of \cite{mo94}. For
simplicity, they just drew straight lines between these bounds to
roughly illustrate the areas with three absorber populations which are
permitted.  It can be seen that there is a substantial region permitted
by both the 10 (large scale) and 0.5 Mpc (small scale) correlation
strength (the region where the vertical and horizontal hatching
overlaps).  The majority of the absorbers have to be uncorrelated with
galaxies in order to produce the weak large scale correlation. The
observed small-scale correlation with galaxies can be produced by a
relatively small admixture of halo absorbers (as little as 10\%),
although the data is also consistent with galaxy halos producing up to
30\% of the observed absorption lines. If none of the absorption is
produced by such halos, i.e. if there are no absorbers physically
associated with a galaxy in the sample with measured redshifts, then
the observed correlations are marginally consistent with a 50:50 mix of
random and galaxy-like absorbers.

\cite{morris94} took these results, and showed that a
mixture with a dominant population of randomly distributed absorbers,
along with an admixture of material tidally stripped from galaxies,
both matched the observations, and made physical sense. The 21~cm image
of the streamers of HI in the M81 group published by \cite{yun94}
gives a visually compelling illustration of the effects of tidal
interactions on gas in galaxies. In the meantime, Smooth Particle
Hydrodynamic (SPH) and other numerical modelling of the growth of
structure of the universe was approaching the present epoch (e.g.
\cite{hernquist96}, \cite{katz96}). These simulations showed that
low column density absorption could be widely distributed, and can be
caused by: ``filaments of warm gas, caustics in frequency space
produced by converging velocity flows, high density halos of hot
collisionally ionised gas, layers of cool gas sandwiched between
shocks, and modest local undulations in undistinguished regions of the
intergalactic medium'' (\cite{katz96}). I.e., the prediction from SPH
modelling was that the Lyman~$\alpha$ forest was, like most forests, a
very complex eco-system.

\section{Column density dependence of absorber properties} \label{sec-column}

It therefore came as something of a surprise when \cite{lanzetta95}, 
and their accompanying press release, announced that ``...
(the) mysterious clouds of hydrogen in space may actually be vast halos
of gas surrounding galaxies. This conclusion runs contrary to the
longstanding belief that these clouds occur in intergalactic space''
(HST Press release \# STScI-PR95-22). Their survey of 46 galaxies near
the LOS to 6 QSOs observed with the FOS as part of the HST Absorption
line Key Project showed that at least 32$\pm$10\% and possibly as high
as 60$\pm$19\% of the absorbers detected in those LOS were part of the
halos of L* galaxies. A couple of lines of observational evidence
supported this conclusion. First, as shown in their figure~23,
(although one might quibble about the sample definition and effects of
upper limits) there is a correlation between absorber Equivalent Width
(EW) and impact parameter to the nearest galaxy. Secondly, independent
work by \cite{dinshaw95} was showing that absorber `sizes',
determined from observations of close pairs of QSOs with the FOS, were
very large -- several hundred kpc. Also there was some theoretical
support for this idea, as discussed for example in \cite{maloney92},
\cite{salpeter93}, and \cite{salpeter95}. An nice independent 
commentary on this issue is given by \cite{carswell95}.

How can these two results be reconciled? It should first be admitted
that the uncertainties in both papers (\cite{morris93} and \cite{lanzetta95}) 
are large enough that they could well be consistent with
each other. However, it seems more likely to me that, as suggested by
the SPH results of \cite{hernquist96}, the correlation between
absorbers and galaxies has a fairly strong dependence on column
density. The correlation claimed by \cite{lanzetta95} is based on
lines with EW greater than 0.3{\AA}, while the sample studied by \cite{morris93} 
was dominated by lines with EW $\sim$0.03-0.1{\AA}. For
typical line widths, this maps on to column densities greater than
10$^{14}$~cm$^{-2}$ for \cite{lanzetta95}, and from
1-3$\times$10$^{13}$~cm$^{-2}$ for \cite{morris93}.

\cite{stocke95}, \cite{lebrun96}, and \cite{bowen96}
investigate the correlation between EW and impact parameter in their
figures~4, 4 and 3 respectively. \cite{stocke95} used the GHRS to
observe a well chosen sample of Seyfert galaxies known to lie on the
far side of voids in the galaxy distribution. Their main conclusion is
that some absorbers are indeed found within the voids, although there
was evidence that the absorber density is lower in voids than
elsewhere. They also show that the claimed correlation between absorber
EW and impact parameter breaks down for EW less than 0.3{\AA}. \cite{lebrun96} 
and \cite{bowen96} use LOS observed with the FOS to
cast doubt on even the large EW end of the correlation. As is commonly
claimed in such situations ``more obervations are needed''. It would be
very interesting if a large transition in correlation properties could
be shown to occur between the `GHRS' and `FOS' column density ranges.

\section{Work in progress} \label{sec-work}

In this review, for pedagogical reasons, I have neglected to mention
the issue of cloud-cloud correlations, and chosen to focus on
cloud-galaxy correltations instead. The former area is nevertheless one
of great interest for people studying clouds at both high and low
redshift. In a poster at this conference, some recent GHRS results are
presented by \cite{tripp96}, suggesting there is significant clustering
of weak Lyman $\alpha$ absorbers around strong ones at low redshift.

\begin{figure}[ht]
\plotfiddle{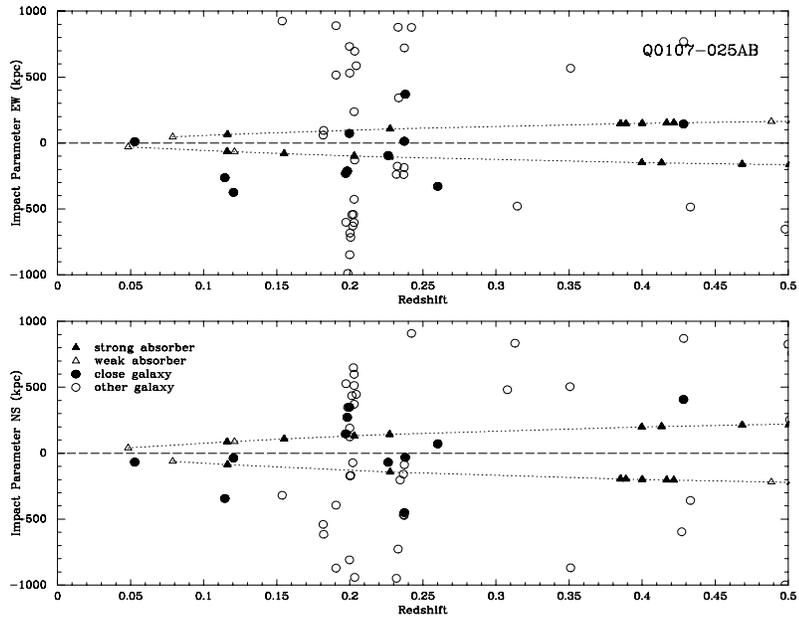}{9cm}{90}{50}{50}{198}{-18}
\caption{Pie-diagrams for the galaxies and absorbers in the LOS to the
QSO pair Q0107-025A,B. \label{fig-morriss6.eps}}
\end{figure}

I would like to conclude this review with one piece of work in progress
which might help clarify the strong EW situation. One problem with most
QSO LOS is that one has no idea of the shape of the absorber projected
on the sky, let alone in three dimensions. In
figure~\ref{fig-morriss6.eps},
I show results from redshift surveys around the QSO pair Q0107-025A,B.
The low-z absorbers were found in GHRS observations were taken with the
G140L grating. Because of the comparative faintness of the QSOs, higher
dispersion observations were not possible. This figure includes 32
galaxy redshifts from CFHT and also redshifts from the Palomar 5m. Of
particular interest are the Lyman $\alpha$ absorbers seen in both lines
of sight. One near z$\sim$0.23 seems to be associated with a bright
early type galaxy near both lines of sight. In contrast, any galaxy
associated with the absorption near z$\sim$0.4, which includes a Lyman
limit system in the `B' line of sight, remains undiscovered. With this
data, we can now at least begin to map out the projected shapes of
absorbers.

\acknowledgments

I would like to thank to Ray Weymann for years of good advice and
interesting ideas on this topic.


\begin{thebibliography}{}
\bibitem[Bahcall et al. 1991]{bahcall91} Bahcall, J. N., Jannuzi, B. T., Schneider, D. P., Hartig, G. F., Bohlin, R. and Junkkarinen, V., 1991, \apjl, 377, L5
\bibitem[Bahcall et al. 1993]{bahcall93} Bahcall et al. 1993, \apjs, 87, 1
\bibitem[Bahcall et al. 1996]{bahcall96} Bahcall et al. 1996, \apj, 457, 19
\bibitem[Bechtold 1994]{bechtold94} Bechtold, J. 1994, \apjs, 91, 1
\bibitem[Bland-Hawthorn et al. 1994]{bland94} Bland-Hawthorn, J., Taylor, K., Veilleux, S. and Shopbell, P. L. 1994, \apjl, 437, L95
\bibitem[Bowen, Blades and Pettini 1996]{bowen96} Bowen, D. V., Blades, J. C. and Pettini, M. 1996, \apj, 464 141
\bibitem[Brandt et al. 1993]{brandt93} Brandt, J. C. et al. 1993, \aj, 105, 831
\bibitem[Carswell 1988]{carswell88} Carswell, R. F., 1988, in ``QSO Absorption Lines: probing the Universe'', STScI Symposium Series \#2, eds. C. J. Blades, D. Turnshek and C. A. Norman, (Cambridge: Cambridge University Press), 91
\bibitem[Carswell 1995]{carswell95} Carswell, R. F. 1995, {\rm Nature}, 374, 500
\bibitem[Cowie et al. 1995]{cowie95} Cowie, L., Songaila, A., Kim, T. and Hu, E. 1995, \aj, 109, 1522
\bibitem[Dinshaw et al. 1995]{dinshaw95} Dinshaw, N., Foltz, C. B., Impey, C. D., Weymann, R. J. and Morris, S. L. 1995, Nature, 373, 223
\bibitem[Hernquist et al. 1996]{hernquist96} Hernquist, L., Katz, N., Weinberg, D. H. and Miralda-Escude, J. 1996, \apjl, 457, L51
\bibitem[Katz et al. 1996]{katz96} Katz, N., Weinberg, D. H., Hernquist, L. and Miralda-Escude, J. 1996, \apjl, 457, L57
\bibitem[Hunstead 1988]{hunstead88} Hunstead, R. W., 1988, in ``QSO Absorption Lines: probing the Universe'', STScI Symposium Series \#2, eds. C. J. Blades, D. Turnshek and C. A. Norman, (Cambridge: Cambridge University Press), 71
\bibitem[Lanzetta et al. 1995]{lanzetta95} Lanzetta, K. M., Bowen, D. B., Tytler, D. and Webb, J. K. 1995, \apj, 442, 538
\bibitem[Le Brun et al. 1996]{lebrun96} Le Brun, V. and Bergeron, J. and Boisse, P. 1996, \aa, 306, 691
\bibitem[Lu, Wolfe and Turnshek 1991]{lu91} Lu, L., Wolfe, A. M. and Turnshek, D. A. 1991, \apj, 367, 19
\bibitem[Lu et al. 1996]{lu96} Lu, L., Sargent, W., Womble, D. and Takada-Hidai, M. 1996, \apj, In Press, astro-ph/9606033
\bibitem[Maloney 1992]{maloney92} Maloney, P. 1992, \apj, 398, 89
\bibitem[Mo and Morris 1994]{mo94} Mo, H. J. and Morris, S. L., 1994, \mnras, 269, 52, 
\bibitem[Morris et al. 1991]{morris91} Morris, S. L., Weymann, R. J., Savage, B. D., Gilliland, R. L. 1991, \apj, 377, L21
\bibitem[Morris et al. 1993]{morris93} Morris, S. L., Weymann, R. J., Dressler, A., McCarthy, P. J., Smith, B. A., Terrile, R. J., Giovanelli, R. and Irwin, M. 1993, \apj, 419, 524
\bibitem[Morris and van den Bergh 1994]{morris94} Morris, S. L. and van den Bergh, S. 1994, \apj, 427, 696
\bibitem[Rauch and Haehnelt 1995]{rauch95a} Rauch, M. and Haehnelt, M. G. 1995, \mnras, 275, L76
\bibitem[Rauch, Weymann and Morris 1995]{rauch95b} Rauch, M., Weymann, R. J. and Morris, S. L., 1995, \apj, 458, 518
\bibitem[Salpeter 1993]{salpeter93} Salpeter, E. E. 1993, \aj, 106, 1265
\bibitem[Salpeter and Hoffman 1995]{salpeter95} Salpeter, E. E. and Hoffman, G. L. 1995, \apj, 441, 51
\bibitem[Savage et al. 1993]{savage93} Savage, B. D., L. Lu, Weymann, R. J., Morris, S. L. and Gilliland, R. L.  1993, \apj,  404, 124
\bibitem[Songaila and Cowie 1996]{songaila96} Songaila, A.  and Cowie, L. L. 1996, \aj, 112, 335
\bibitem[Stocke et al. 1995]{stocke95} Stocke, J. T., Shull, J. M., Penton, S., Donahue, M. and Carilli, C. 1995, \apj, 451, 24
\bibitem[Tripp, Lu and Savage 1996]{tripp96} Tripp, T. M., Lu, L. and Savage, B. D. 1996, This Proceedings
\bibitem[Van Gorkom et al. 1993]{vangorkom93} Van Gorkom, J. H., Bahcall, J. N., Jannuzi, B. T. and Schneider, D. P. 1993, \aj, 106, 2213 
\bibitem[Weymann et al. 1995]{weymann95} Weymann, R. J., Rauche, M., Williams, R., Morris, S. L. and Heap, S., 1995, \apj, 438, 650
\bibitem[Williams and Schommer 1993]{williams93} Williams, T. B. and Schommer, R. A. 1993, \apjl, 4 
\bibitem[Yun, Ho and Lo 1994]{yun94} Yun, M. S., Ho, P. T. P. and Lo, K. Y. 1994, {\rm Nature}, 372, 530
\end{thebibliography}
\end{document}